\documentclass[twocolumn,aps,floatfix]{revtex4}
\usepackage{amssymb} \usepackage{graphicx} \usepackage{amsmath}
\usepackage[T1]{fontenc} \usepackage{pstricks} \usepackage{subfigure}
\usepackage[colorlinks=true,citecolor=blue,urlcolor=red]{hyperref}
\usepackage[normalem]{ulem}
\usepackage{hyperref}
\usepackage{doi}

\begin{document}
\title{Nonzero temperature dynamics of a repulsive two-component Fermi gas}

\author{Jaros{\l}aw Ryszkiewicz,$\,^1$ Miros{\l}aw Brewczyk,$\,^1$ and Tomasz Karpiuk$\,^1$}

\affiliation{\mbox{$^1$ Wydzia{\l} Fizyki, Uniwersytet w Bia{\l}ymstoku,  ul. K. Cio{\l}kowskiego 1L, 15-245 Bia{\l}ystok, Poland } }

\date{\today}

\begin{abstract}

We study spin-dipole oscillations of a binary fermionic mixture at nonzero temperatures. We apply the atomic-orbital method combined with the Monte Carlo technique based sampling to probe finite temperatures. Our results agree quantitatively with recent experiment, G. Valtolina et {\it al}., Nat. Phys. {\bf 13}, 704 (2017), showing the appearance of the ferromagnetic phase at stronger repulsion between components when the temperature is increased.

\end{abstract}

\maketitle 

The long-standing picture of itinerant ferromagnetism posed by E. Stoner in his 1933 pioneering work \cite{Stoner33} is now being tested experimentally. As it was predicted in \cite{Stoner33}, free (i.e. not localized) electrons become ferromagnetic when a short-range screened Coulomb repulsion between opposite spin particles gets strong enough to overcome the Fermi pressure. To verify the original Stoner's idea one needs a system free of all these, typical for solids, beyond short-range repulsion effects \cite{Brando16}. Cold fermionic atoms seem to be the solution. Indeed, by using fermionic atoms one can study the interplay between fermionic quantum pressure and the short range interatomic repulsive interactions \cite{DeMarco02,Du08,Jo09,Sommer11,Sanner12,Lee12,Valtolina17}. Hence, the question whether a strong repulsion between atoms can overcome the effect of a pressure, which tends to dispose atoms in the whole available space, and lead to formation of spatial domains can be addressed.

Attempts have been already undertaken to verify experimentally the Stoner's idea of itinerant ferromagnetism \cite{Jo09,Sommer11,Sanner12,Lee12,Valtolina17}. The enthusiasm for the use of cold fermionic atoms, however, quickly weakened. The reason is that the repulsive Fermi gas is, in fact, a metastable state as it corresponds to the excited many-body energy branch \cite{Chin10}. While driving the atoms into the strong repulsive regime, the process of forming bound states becomes more competing and the bound molecular states are formed. These bound states are reached through the three-body collisions which become crucially important near the Feshbach resonance (see discussion of the pairing versus ferromagnetic instabilities in a recent experimental work of Ref. \cite{Amico18}).

To avoid difficulties related to pairing effect, in the experiment \cite{Valtolina17} $^6$Li atoms mixture was initially prepared in a state which mimic the ferromagnetic one. For that, both components, held in a cigar-shaped harmonic trap, were first spatially separated by using a magnetic field gradient. When the overlap between two atomic clouds was negligible, the optical repulsive barrier was turned on and the magnetic gradient put off.

Two kinds of experiments were performed with the system prepared in such a state. In the first type the spin dynamics is investigated by sudden switching off the optical barrier. In this case the frequencies of the spin-dipole mode are measured, they are extracted from the time dependent behavior of the relative distance between centers of two spin clouds. This measurement reveals the existence of the critical repulsion. In the weakly interacting regime the spin-dipole frequency decreases with the increase of the intercomponent repulsion. This softening of the spin-dipole mode continues until some critical repulsion strength is reached. For stronger interaction the spin dynamics changes qualitatively, the clouds stop to pass through each other and start to bounce off each other with frequency higher than the trap frequency. The spin-dipole mode is studied for two temperatures only, much below the Fermi temperature.

In the second experiment the stability of two spin domains is investigated. For that the separating optical barrier is removed adiabatically and the spin diffusion effect is observed below and above the critical interaction strength. No detectable spin-dipole oscillations are excited in this case. Instead, the magnetization of the sample is measured and the time windows of constant magnetization are observed above the critical repulsion. The duration of time windows is, however, finite because of relaxation of the system to the lower-lying energy branch. The periods of constant magnetization start to appear at the interaction strength very close to that at which the spin-dipole frequency, after continues decreasing, shows a sudden jump to higher values. Thus, both kinds of experiments support the existence of the critical repulsion, above which the paramagnetic phase becomes unstable in favor of the ferromagnetic one \cite{Stoner33,Pilati10,Massignan11,Chang11,Pekker11,Massignan14,Trappe16}.

To model the experiment of \cite{Valtolina17} we use the Hartree-Fock approximation (already applied to study various mixtures of cold atoms \cite{Karpiuk04a,Karpiuk04b,Karpiuk06,Grochowski17}), i.e. we treat all fermionic atoms individually by assigning single particle wave functions to each of them. These wave functions, so called atomic orbitals, depend both on spatial and spin degrees of freedom. Then the many-body wave function of the system of $N$ indistinguishable fermionic atoms, in its simplest form, can be approximated by the single Slater determinant

\begin{eqnarray}
&&\Psi ({\bf x}_1,...,{\bf x}_{N})
= \frac{1}{\sqrt{N!}} \left |
\begin{array}{lllll}
\phi_1({\bf x}_1) & . & . & . & \phi_1({\bf x}_{N}) \\
\phantom{aa}. &  &  &  & \phantom{aa}. \\
\phantom{aa}. &  &  &  & \phantom{aa}. \\
\phantom{aa}. &  &  &  & \phantom{aa}. \\
\phi_{N}({\bf x}_1) & . & . & . & \phi_{N}({\bf x}_{N})
\end{array}
\right |  .   \nonumber  \\
\label{Slater}
\end{eqnarray}
The coordinates ${\bf x}_n$ ($n=1,...,N$) of atoms include spatial and spin variables and $\phi_n({\bf x})$ $(n=1,...,N)$ are the orthonormal spin-orbitals. Here we consider a two-component Fermi gas, then the spin-dependent part of the spin-orbitals is twofold. As in experiment \cite{Valtolina17}, we assume a balanced mixture of two spin states.

Fermions occupying the same spin state do not interact. The only considered interaction in the system is the repulsion between different spin atoms. At low temperatures, on many occasions, it is well described by the contact potential with the coupling constant $g$, related to the $s$-wave scattering length $a$ via $g=4\pi \hbar^2 a/m$. For such spin-dependent interactions, the time-dependent Hartree-Fock equations for the spatial parts of the spin-orbitals, $\phi_n^{+}({\bf r},t)$ and $\phi_n^{-}({\bf r},t)$, can be written as

\begin{eqnarray}
i\hbar\, \frac{\partial}{\partial t}\,  \phi_n^{\pm} ({\bf r},t) = {\cal H}_{sp}\,  \phi_n^{\pm} ({\bf r},t)
\label{HFeq}
\end{eqnarray}
for $n=1,...,N/2$. The effective single-particle Hamiltonian is given by ${\cal H}_{sp}=-\frac{\hbar^2}{2 m} \nabla^2 + V_{tr}({\bf r}) + g\, n_{\mp} ({\bf r},t)$ with atomic densities of components (normalized to the number of atoms in each component), $n_{+} ({\bf r},t)$ and $n_{-} ({\bf r},t)$, defined as $n_{\pm} ({\bf r},t)=\sum_{n=1}^{N/2} |\phi_n^{\pm} ({\bf r},t)|^2$.

However, to retrieve quantitatively the results of experiment \cite{Valtolina17} an oversimplified description of particles' interactions should be improved. One needs to include the many-body correlations raised by the interactions. This can be done by modifying the many-body wave function (\ref{Slater}) by including the Jastrow correlation terms. The quantum Monte Carlo variational calculations with a trial wave function of the Jastrow-Slater form have been already successfully applied to the mixtures of degenerate gases \cite{Pilati10,Bertaina13}. Note, however, that a much simpler approach, the lowest order constrained variational method \cite{Pandharipande71,Pandharipande73}, exists and can be safely used to get a good approximation to the results of quantum Monte Carlo calculations \cite{Cowell02,Yu11}. The time-dependent extension of this simplified approach is accessible as well.

The other way to include correlations is to introduce effective interactions. It was already proposed in \cite{Grochowski17}, where the zero temperature spin-dipole oscillations of two-component Fermi gas were investigated. The idea is to renormalize the coupling parameter in the two-particle contact potential in a way to be able to fulfill the well known low-density expansion of an energy in the dimensionless parameter $k_F a$, where $k_F$ is the Fermi wave number. For two-component Fermi system this energy expansion, with up to third order terms, reads \cite{Fetter,Baker71}
\begin{eqnarray}
\frac{E}{N \varepsilon_F} &=&
\frac{3}{5} + \frac{2}{3\pi} (k_F a) + \frac{4}{35\pi^2} (11-2\ln{2})\, (k_F a)^2 \nonumber  \\
&+& 0.23\, (k_F a)^3 + ...   \,,
\label{expansion}
\end{eqnarray}
where $\varepsilon_F$ is the Fermi energy. The term proportional to $k_F a$ is just the mean-field expression for the interaction energy. The next term, derived first by Huang and Yang \cite{Huang57} and Lee and Yang \cite{Lee57}, takes into account the modification of the intermediate states due to the Pauli exclusion principle. The last one results from the three-body correlations \cite{DeDominicis57} and, in principle, depends on $s$- and $p$-wave scattering lengths as well as on the $s$-wave effective range. The value given in (\ref{expansion}) is just the one calculated for the hard-sphere potential case. In the case of attractive interactions, on the Bardeen-Cooper-Schrieffer side of the so-called BCS-BEC crossover, the formula corresponding to (\ref{expansion}) was already verified experimentally, see Ref. \cite{Navon10}. To follow the formula (\ref{expansion}) one has to renormalize the coupling constant locally \cite{Stecher07}, which effectively results in the replacement of $g n_\pm$ term in Eqs. (\ref{HFeq}) by $g n_{\pm} + A (4/3\, n_{\mp}^{1/3}\, n_{\pm} + n_{\pm}^{4/3}) + B (5/3\, n_{\mp}^{2/3}\, n_{\pm} + n_{\pm}^{5/3})$, where $A=3 g a (6\pi^2)^{1/3} (11-2\ln{2}) /35 \pi$ and $B=3 g a^2 (6\pi^2)^{2/3} \pi/4 \times 0.23$ \cite{Grochowski17}.

To extend our analysis by including temperature effects on spin-dipole oscillations of a two-component Fermi gas, we allow to populate single-particle states of energies higher than the Fermi energy. It is done with the help of the Monte Carlo technique based sampling. 
Although the Eqs. (\ref{HFeq}) include the mean-field term representing the repulsion between different spin atoms, the initial state of the two-component Fermi gas is actually free of interactions. This is because initially both spin-up and spin-down atomic clouds are spatially separated, just like in the experiment \cite{Valtolina17}. Since cold polarized fermionic atoms do not interact, it is justified to generate the grand canonical ensemble of many-body states of two-component Fermi gas, corresponding to the initial state, assuming the energy levels and their populations as for the ideal gas. Then the probability of populating a one-particle state $\phi_n$ of energy $\varepsilon_n$ is given by the Fermi-Dirac distribution $p_n=(\exp[\beta (\varepsilon_n - \mu)]+1)^{-1}$, where $\beta$ determines the bath temperature ($\beta = 1/ k_B T$) and $\mu$ is the chemical potential. 
We generate, according to this probability, a number of many-particle configurations (states) for both components for each temperature. Technically speaking, a many-body state is built by going through the set of considered single-particle states and accepting each of them with the probability $p_n$. For that for each single-particle state $\phi_n$ a random number $r$ from the interval $[0,1]$ is drawn and compared with the probability $p_n$. The single-particle state is accepted provided $r<p_n$. The numbers of atoms in each component slightly differ from one configuration to the other but on average they are the same provided the number of many-body states drawn is large enough.

To follow the above prescription we need to fix the value of the chemical potential for each fermionic component. The chemical potential and the temperature are two control parameters in the grand canonical ensemble. They can be related to the average number of atoms and the average energy in the ensemble. To find the chemical potential corresponding to the given average number of atoms, $\langle N \rangle$, at a given temperature one must solve the equation $\sum_{n=1}^{n_{max}} (\exp[\beta\, (\varepsilon_n - \mu)]+1)^{-1} = \langle N \rangle$, the left-hand side of which is just the sum of average occupations of all considered single-particle states. Throughout this work we have $\langle N \rangle =N/2=24$ for each component and the number of single-particle states taken into account is $n_{max} \approx 4\times 10^3$. The chemical potential found in this way is close to the one given by the low temperature Sommerfeld expansion for a gas in a harmonic trap $\mu=\varepsilon_F[1-(\pi^2/3) (k_B T/\varepsilon_F)^2]$, for temperatures up to $0.45\, T_F$ ($T_F$ is the Fermi temperature). The above Sommerfeld expansion turned out to be correct for the ideal Fermi gas trapped in a harmonic potential for temperatures even up to $0.55\, T_F$ \cite{Rokhsar97}. 

\begin{figure}[t!hb] 
\includegraphics[width=7.9cm]{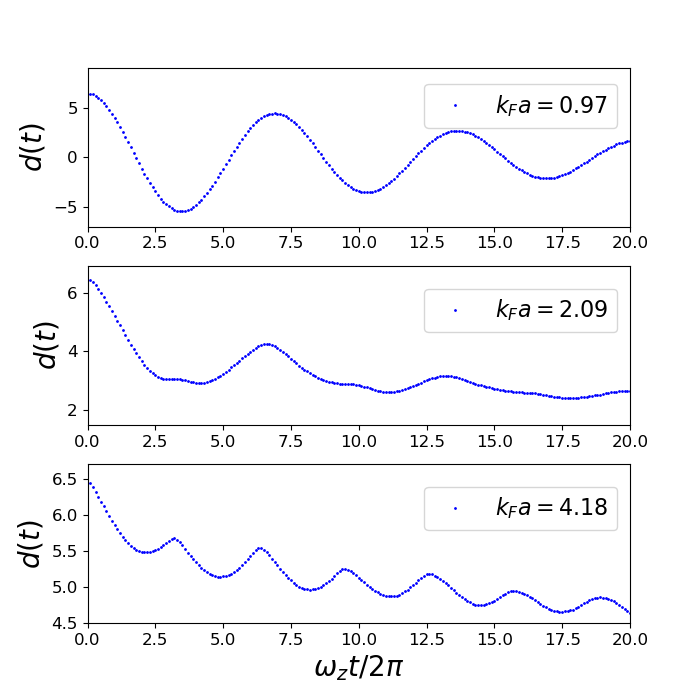}
\caption{Spin-dipole oscillations of atomic clouds for increased interaction strength, $k_F a$, at the temperature $T/T_F=0.4$. The relative distance $d(t)$ between the centers of mass of two spin clouds is shown as a function of time. Each value $d(t)$ is an average over $10$ configurations. The number of atoms is $N/2=24$. }  
\label{oscillations}
\end{figure}

\begin{figure}[h!tb] 
\includegraphics[width=8.2cm]{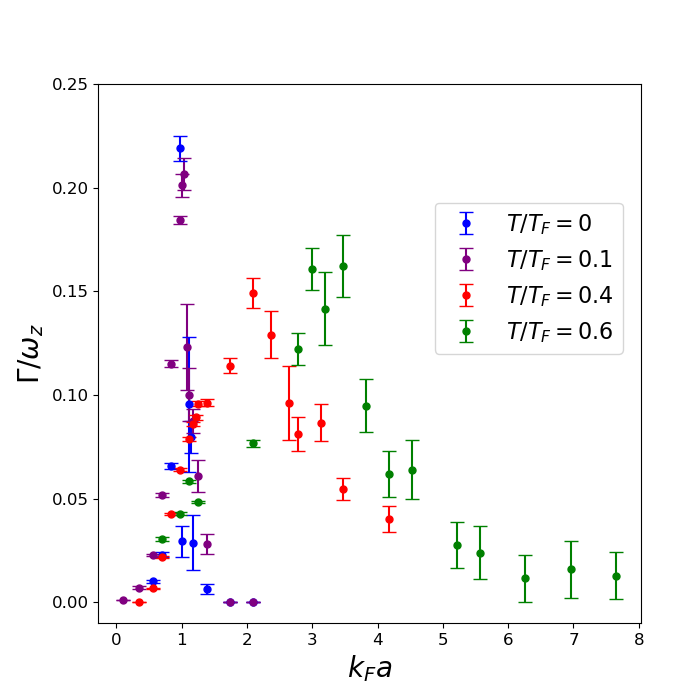}
\caption{Damping rates for different repulsion strengths and a number of temperatures. As in experiment \cite{Valtolina17}, we find a strong increase of the damping of the spin-dipole oscillations while approaching the critical repulsion. After crossing the critical region the damping of oscillations gets again low.    }
\label{damping}
\end{figure}

In this way, having given the temperature $T$ and the average number of atoms $N/2$ in each component, we sample, with the help of the Fermi-Dirac distribution, the many-particle states space for a two-spin Fermi gas. Then, for each many-particle state we abruptly remove the barrier separating components and start dynamics of two atomic clouds. For that we numerically solve the set of Eqs. (\ref{HFeq}) \cite{Gawryluk18} for different temperatures and the interaction strengths $k_F a$. Here, $k_F$ is the Fermi wave number equal to $k_F=(24 N)^{1/6}/a_{ho}$, where $a_{ho}=(\hbar/m\, \omega_{ho})^{1/2}$ and $\omega_{ho}$ is the geometrical average of trapping frequencies. As in experiment \cite{Valtolina17}, the axial and radial trap frequencies are equal to $\omega_z=2\pi \times 21\,$Hz and $\omega_{\perp}=2\pi \times 265\,$Hz. We monitor the separation $d(t)$ between the centers of mass of two atomic clouds as a function of time. The distance $d(t)$ differs, in general, between configurations, therefor the results are averaged over $10$ samples. In Fig. \ref{oscillations} we plot averaged $d(t)$ for the temperature $T/T_F =0.4$ and for different interaction strengths. We depict three qualitatively distinct regimes of dynamics of the system. For low and high enough $k_F a$ (upper and lower frames, respectively) $d(t)$ clearly oscillates. In the crossing regime the oscillations are strongly dumped (middle frame and Fig. \ref{damping}).

\begin{figure}[h!tb] 
\includegraphics[width=8.0cm]{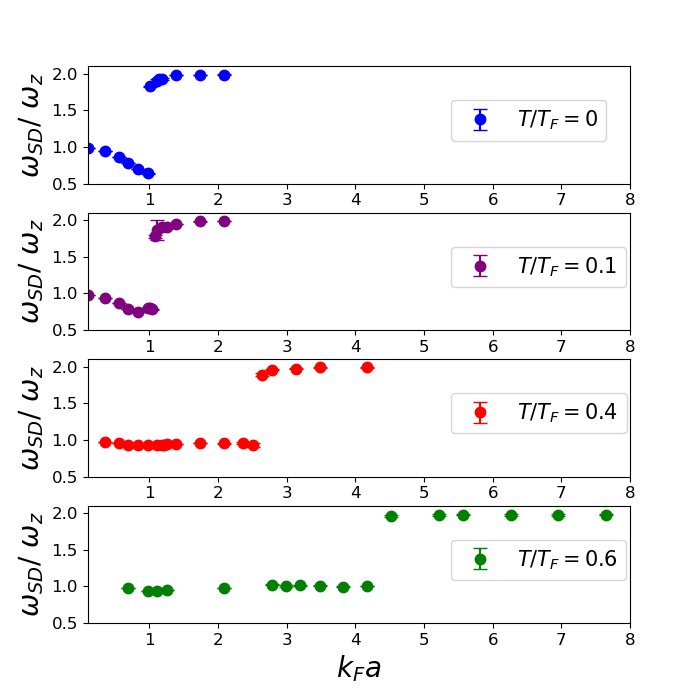}
\caption{Frequencies of the spin-dipole mode of a repulsive two-component Fermi gas plotted as a function of $k_F a$ for different temperatures, increasing from top to bottom. The higher temperature impedes the formation of the ferromagnetic phase, therefor the separation of the components occurs for larger $k_F a$. }  
\label{frequencies}
\end{figure}

Analyzing data as in Fig. \ref{oscillations}, we fit the averaged distance $d(t)$ to $c_0 + c_1 t + c_2\, e^{-\Gamma t} \sin{(\omega_{SD} t + \varphi)}$. Figs. \ref{frequencies} and \ref{damping} summarize our results with respect to the fitting parameters $\omega_{SD}$ and $\Gamma$. For weak repulsion two atomic clouds behave as miscible fluids. In the limit of noninteracting clouds they oscillate with the frequency equal to the axial trap frequency. When the repulsion increases the frequency gets lower, down to $0.5\, \omega_z$ at zero temperature \cite{Grochowski17}. This, so called mode-softening effect is diminished for higher temperatures, see Fig. \ref{frequencies}. When the strength of the repulsion is increased further we observe the qualitative change in the response of the system. Two atomic clouds become immiscible. For temperatures up to $T/T_F=0.4$  atomic clouds oscillate with the frequency below $2\, \omega_z$ just after crossing the critical region and close to twice the axial frequency for larger $k_F a$. The critical value of $k_F a$ is shifted up for higher temperatures. This can be understood based on the Stoner's model of itinerant ferromagnetism. In this model the repulsion between fermions counteracts fermionic quantum pressure. Since the Fermi pressure of an ideal gas gets larger with the temperature \cite{Huang}, the stronger repulsion is needed to overcome the pressure. The same reason causes the decline of the effect of mode-softening for higher temperatures.

Results presented in Fig. \ref{damping}, showing damping rates of the spin-dipole oscillations, are in agreement with experimental observations \cite{Valtolina17}. Below the critical repulsion, while the frequency $\omega_{SD}$ decreases, the damping of the oscillations strongly increases. Beyond the critical region, when the spin-dipole mode frequency $\omega_{SD}$ jumps to the value of $2\, \omega_z$, the damping is significantly reduced. This behavior holds for all considered temperatures.

\begin{figure}[h!tb] 
\includegraphics[width=7.8cm]{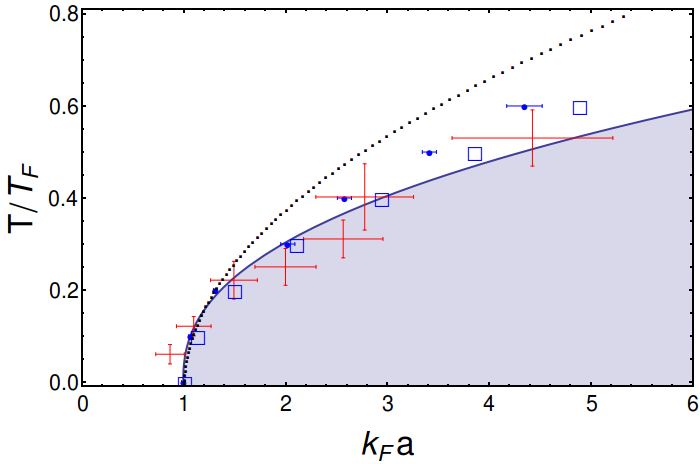}
\caption{Critical value of the repulsive interaction strength, $(k_F a)_{cr}$, at a given temperature obtained from the analysis of the spin-dipole mode as in Fig. \ref{frequencies} (blue bullets, with additional horizontal bars showing the extent of dispersion). Beyond $(k_F a)_{cr}$ the system remains in the ferromagnetic state (gray area). The red crosses are the experimental data taken from Ref. \cite{Valtolina17} (Fig. 3d), related to the measurement of a diffusion of two spin domains. The blue bullets come from numerics, after averaging over $10$ configurations for each temperature. The dotted line is plotted based on the lowest-order Sommerfeld expansion, whereas for the case of blue squares the energy is calculated numerically. The solid line is a power-law fit to the numerical points for temperatures $T/T_F < 0.4$. The number of atoms is $N=48$.   } 
\label{TvkFa}
\end{figure}

Finally, we gather our data in Fig. \ref{TvkFa}, where we plot the values of the critical repulsion $(k_F a)_{cr}$ for studied temperatures. The numerical results are denoted by blue bullets and come from the analysis of oscillations of the spin-dipole modes as in Fig. \ref{frequencies}. We also show the experimental data, red crosses, taken from Ref. \cite{Valtolina17}, which are collected based on the measurement of the stability of the ferromagnetic state against the spin diffusion. According to \cite{Valtolina17}, metastable ferromagnetic state appears for the repulsion $k_F a$ very close to the one at which the frequency of the spin-dipole mode exhibits a sudden jump to high values. Indeed, our numerical results are in agreement with experimental data, even though they are obtained for a system much smaller than studied in \cite{Valtolina17}. Such behavior, i.e. the independence of $(k_F a)_{cr}$ on the number of particle was already recognized in Ref. \cite{Grochowski17} for a zero temperature case.

To understand this universal behavior for a gas at nonzero temperatures we adopt the Stoner's picture of itinerant ferromagnetism. To find the critical value of repulsive interactions we compare the kinetic energy of the gas to the interaction energy \cite{Zwerger09}. For a uniform system at zero temperature, within the Thomas-Fermi approximation, it gives $(k_F a)_{cr} = \pi/2$. Including temperature, through the lowest order Sommerfeld expansion for the internal energy \cite{Huang}, into considerations one has $(k_F a)_{cr} = \pi/2\, [1+5\pi^2 /12\, (T/T_F)^2]$. Two corrections to this formula are needed. First, we know from the experiment \cite{Valtolina17} and a number of theoretical papers \cite{Conduit09,Pilati10,He16,Grochowski17} that at zero temperature the critical repulsion is, in fact, smaller and closer to one. In Ref. \cite{Grochowski17}, to get the correct value, we renormalized the coupling constant in the interparticle interactions consulting correlations in this way. Second, the temperature dependent factor should correspond rather to the harmonic potential case. Therefor, we have $(k_F a)_{cr} \approx [1+2\pi^2 /3\, (T/T_F)^2]$ and the border between paramagnetic and ferromagnetic phases is denoted in Fig. \ref{TvkFa} as a dotted line. An agreement with numerics holds only for low temperatures as the Sommerfeld expansion does apply in this range. We can improve an agreement by calculating the energy of the system according to the grand canonical ensemble rules as $E(T)=\sum_{n=1}^{n_{max}} \varepsilon_n (\exp[\beta\, (\varepsilon_n - \mu(T))]+1)^{-1}$. The results, i.e. $T/T_F$ as a function of $(k_F a)_{cr}=E(T)/E(T=0)$ are visible in Fig. \ref{TvkFa} as blue squares. Now, an agreement remains at the quantitative level for all temperatures.

One might be surprised that the above formula for the energy, $E(T)$, of the system as it was the ideal gas works well. This is because at the critical value of $k_F a$ two atomic components become separated. Of course, this separation is not as perfect as it was initially, at time equal to zero. There exists some domain wall in between components, where atoms of both spins are mixed and repulsively interacting, to support the separation. Also, at $(k_F a)_{cr}$ the gas is still unpolarized on the perimeter (the overlapping part vanishes only for strong enough repulsion). Surely, the presence of the domain wall modifies the single particle energies, $\varepsilon_n$, for larger $n$. This change gets increased with temperature (or the interaction strength) since then the effective interspecies barrier decreases and becomes broader. Therefore, for higher temperatures the approximate results (blue squares in Fig. \ref{TvkFa}) departure from the numerical ones (blue bullets). Nevertheless, the overall agreement remains good.

In summary, we have studied dynamics of a repulsive two-component Fermi gas at nonzero temperatures. We utilize the atomic-orbital method and apply the Monte Carlo sampling to probe many-particle states due to finite  temperatures. We find a quantitative agreement with experimental results of \cite{Valtolina17} showing the dependence of the critical repulsion $k_F a$, indicating the transition to the ferromagnetic phase, on the temperature. The critical strength $k_F a$ increases with temperature in accordance with the Stoner's picture of itinerant ferromagnetism.

\acknowledgments  
We are grateful to K. Rz\c{a}\.zewski and P. Grochowski for discussions. The authors acknowledge support from the (Polish) National Science Center Grant No. 2018/29/B/ST2/01308. Part of the results were obtained using computers at the Computer Center of University of Bialystok.


\end{document}